# Comparative Spectra of Oxygen-Rich vs. Carbon-Rich Circumstellar Shells: VY Canis Majoris and IRC+10216 at 215-285 GHz

E. D. Tenenbaum [1,2,5]

J. L. Dodd [1,2]

S. N. Milam[3]

N. J. Woolf [1]

and

L. M. Ziurys [1,2,4]

[1] Department of Astronomy and Steward Observatory, University of Arizona, 933 N. Cherry Ave. Tucson, AZ USA 85721. tenenbaum@strw.leidenuniv.nl, jldodd@email.arizona.edu, lziurys@email.arizona.edu, nwoolf@as.arizona.edu
[2] Department of Chemistry, University of Arizona
[3] NASA Goddard Space Flight Center, Astrochemistry Laboratory, Code 691, Greenbelt, MD 20771, USA, Stefanie.N.Milam@nasa.gov
[4] Arizona Radio Observatory
[5] current address: Sackler Laboratory for Astrophysics, Leiden Observatory, Leiden University, P.O. Box 9513, NL-2300 RA Leiden, The Netherlands

**Abstract:**

A sensitive (1σ rms at 1 MHz resolution ~3 mK) 1 mm spectral line survey (214.5-285.5 GHz) of VY Canis Majoris (VY CMa) and IRC+10216 has been conducted to compare the chemistries of oxygen and carbon-rich circumstellar envelopes. This study was carried out using the Submillimeter Telescope (SMT) of the Arizona Radio Observatory (ARO) with a new ALMA-type receiver. This survey is the first to chemically characterize an O-rich circumstellar shell at millimeter wavelengths. In VY CMa, 128 emission features were detected arising from 18 different molecules, and in IRC+10216, 720 lines were observed, assigned to 32 different species. The 1 mm spectrum of VY CMa is dominated by $SO_2$ and SiS; in IRC +10216, $C_4H$ and $SiC_2$ are the most recurrent species. Ten molecules were common to both sources: CO, SiS, SiO, CS, CN, HCN, HNC, NaCl, PN, and $HCO^+$. Sulfur plays an important role in VY CMa, but saturated/unsaturated carbon dominates the molecular content of IRC+10216, producing $CH_2NH$, for example. Although the molecular complexity of IRC+10216 is greater, VY CMa supports a unique "inorganic" chemistry leading to the oxides PO, AlO, and AlOH. Only diatomic and triatomic compounds were observed in VY CMa, while species with 4 or more atoms are common in IRC+10216, reflecting carbon's ability to form strong multiple bonds, unlike oxygen. In VY CMa, a new water maser ($v_2$=2) has been found, as well as vibrationally-excited NaCl. Toward IRC+10216, vibrationally-excited CCH was detected for the first time.

## 1. Introduction

The chemistry of carbon-rich circumstellar shells of evolved stars has been the subject of investigation since the early days of millimeter astronomy (e.g. Wilson et al. 1971). One approach to elucidating the chemical properties of such objects is to continuously measure their spectra across a wide frequency range, namely, a spectral line survey. A number of such surveys have been carried out for C-rich envelopes, in particular that of IRC+10216, the best studied AGB star of this type, which lies at a distance of 150 pc and has a mass loss rate of ~$3\times10^{-5}$ $M_{\odot}yr^{-1}$ (Agúndez & Cernicharo 2006). In total, seven band scans of IRC+10216 have been performed (e.g Avery et al. 1992, Kawaguchi et al. 1995, Cernicharo et al. 2000), including the very latest 1 and 2 mm survey by He et al. (2008). More recently, a 80-276 GHz spectral survey has also been conducted of the post-AGB star CRL 618 (Pardo et al. 2007). These studies, along with ISO and others, have helped to demonstrate the presence of 71 different chemical compounds in IRC+10216, and about 20 in CRL 618. Carbon-rich circumstellar shells thus remain some of the most chemically complex objects in the Galaxy.

It is estimated that perhaps 50% of evolved stars have oxygen-rich, as opposed to carbon-rich, circumstellar envelopes (Kwok 2002). Observations have shown that on the Red Giant/Supergiant branches, stars are oxygen-rich (O > C), as well as many on the AGB (e.g. Milam et al. 2009). It is thought that the "third dredge-up" on the thermal pulsing AGB branch converts an oxygen-rich star to a carbon-rich one, hence the appearance of objects such as IRC+10216 (Herwig 2005). Nonetheless, the regular occurrence of O-rich circumstellar shells would suggest that they are chemically relevant and theorists have considered them in their modeling (Willacy & Millar 1997). Yet, few observational studies have been carried out for these objects. Furthermore, only a limited number of molecular species have been identified in O-rich stars (Omont et al. 1993; Olofsson et al. 1991); consequently, it has been speculated that such objects lack chemical complexity (Olofsson 2005).

In order to systematically evaluate the chemistry of oxygen-rich circumstellar shells, and directly compare them to their C-rich analogs, we have conducted a 1 mm (215–285 GHz) spectral line survey of both VY Canis Majoris (VY CMa) and IRC+10216, using the Sub-millimeter Telescope (SMT) of the Arizona Radio Observatory (ARO). VY CMa is an oxygen-rich supergiant star with a massive envelope, at a distance of 1.1 kpc and a mass loss rate of $\sim 3\times10^{-4}$ $M_{\odot} yr^{-1}$ (Smith et al. 2001). This study was motivated in part by the advances in receiver technology resulting from the Atacama Large Millimeter Array (ALMA). The survey was conducted over a 3 year period using an ALMA-type sideband-separating receiver with exceptional noise temperatures, baseline stability, and image rejection. It is the most sensitive survey to date, with typical rms noise levels of 3 mK (1 MHz resolution), and sometimes better, across the entire 70 GHz band. The complete data set is being published in another paper (Tenenbaum et al. Paper II), and an in-depth chemical abundance analysis is underway (Dodd et al., Paper III). Here we present an overview of the relative chemical compositions of VY CMa and IRC+10216, and additional discoveries from the survey.

## 2. Observations

The measurements were conducted 2007 February through 2009 December using the ARO 10 m SMT on Mt. Graham, AZ. The 1 mm receiver used was dual-polarization and employed ALMA Band 6 sideband-separating (SBS) mixers. Image rejection was typically $\geq 15$ dB, and LO shifts were done at all frequencies to establish any image contamination. The temperature scale at the SMT is measured as $T_A^*$, derived by the chopper wheel method, where the radiation temperature is defined as $T_R = T_A^*/\eta_b$, and $\eta_b$ is the main-beam efficiency. Over the 1 mm band, the beam size ranged from $\theta_b$=35-26" and the average beam efficiency was $\eta_b$=0.74. Typical system temperatures were 200-450 K, with receiver temperatures of 60-80 K. Pointing accuracy is estimated to be ~ ±2".

Observations were conducted in beam-switching mode with a ±2′ subreflector throw. The spectra were measured in 1 GHz intervals across the range 214.5-285.5 GHz, using 1 MHz resolution filter banks as backends, typically configured in parallel mode (2x1024 channels). The data were then smoothed to 2 MHz resolution. The average integration time per 1 GHz interval is 8.6 hr, and the total integration time vested in the survey is ~1700 hr. Linear baselines were removed from the data and line parameters were established from the SHELL or GAUSS fitting routines in CLASS. For more detail, see Tenenbaum et al. (Paper II).

## 3. Results

The survey is summarized in Figure 1 and Table 1. Figure 1 displays the entire data sets for both VY CMa and IRC+10216, showing the relative spectral line density. A 1 GHz section of the survey at 267 GHz is shown in more detail. It is obvious from this figure that IRC+10216 contains far more spectral features at the lower intensity levels ($T_A^*$<0.1 K). In fact , 720 individual lines were detected in IRC+10216, as opposed to 130 for VY CMa, a difference of about a factor of 6. Moreover, there are more unidentified lines in the C-rich shell, as the spectrum at 267 GHz illustrates. (Identifications were made on the basis of public catalogs and the spectroscopic literature and checked for internal consistency.) The most abundant molecules in both sources, from preliminary analysis, are CO and HCN, followed by CS and SiS in IRC+10216 and SiO and SiS in VY CMa. A total of 123 U-lines were found in this survey for IRC+10216, but only 13 for VY CMa. Unidentified lines therefore compose about 17% of the total spectral features in IRC+10216 at this wavelength, and 10% in VY CMa. A nearly identical percentage of U-lines was found in the 2 mm survey of IRC+10216 by Cernicharo et al. (2000), using the IRAM 30 m. However, VY CMa is almost a factor of 10 more distant than IRC+10216 (1.1 kpc vs. 150 pc); if it were at 150 pc, the spectral line density in VY CMa would likely significantly increase.

In Table 1, the chemical compounds identified in each source and the number of detected transitions per species are listed. The molecules which are common to both objects appear at the top of each column (gray background). They include six carbon-bearing molecules (CO, HCN, CN, $HCO^+$, HNC, CS), two with silicon (SiO, SiS), and the exotic refractories PN and NaCl. The abundances of these molecules in IRC+10216 are typically an order of magnitude higher than in VY CMa.

In the 1 mm spectrum, 32 different species have been identified in IRC+10216. $SiC_2$ and $C_4H$ in their various isotopologues ($^{30}SiC_2$, $^{29}SiC_2$, $Si^{13}CC$ and $^{13}CCCCH$, $C^{13}CCCH$, $CC^{13}CCH$, and $CCC^{13}CH$) and excited vibrational states ($SiC_2$: $v_3$=1 and 2, $C_4H$: $v^{\ell}_7=1^1$, $2^0$, and $2^2$) are responsible for 36% of the observed lines. Another major contributor, with 47 lines, is SiS in the v=0-4 states and isotopic variants ($^{29}SiS$, $^{30}SiS$, $Si^{34}S$, $Si^{33}S$, $^{30}Si^{34}S$, and $^{29}Si^{34}S$). $SiC_2$, $C_4H$, and SiS account for the majority of lines in IRC+10216 at 2 mm, as well (Cernicharo et al. 2000). NaCN gives rise to 38 total lines, all with weak intensities ($T_A^*$=2-11 mK). The remainder of the features are primarily due to short carbon-chain species and simple refractory molecules with Si, P, Al, and K. Furthermore, a new molecule, $PH_3$, was discovered in this survey (Tenenbaum & Ziurys 2009), and $CH_2NH$ has been identified for the first time in a circumstellar shell (see Section 4.2).

A number of molecules that are present at lower frequencies are notably absent from the 215-285 GHz spectral region, including the carbon chains $HC_9N$, $HC_7N$, $HC_5N$, $C_6H$, $C_5H$, $C_4Si$, $C_3S$, and MgNC. The small rotational constants of these heavy molecules, combined with low excitation conditions, make their emission lines more prominent at 2 and 3 mm.

In VY CMa, 18 different chemical compounds have been identified, almost doubling the known chemical inventory in O-rich shells. For those with only one transition accessible in the 1 mm band (e.g. CS, HNC, $HCO^+$, $H_2S$), confirming lines were measured at 2 and 3 mm (see Ziurys et al. 2007, 2009). Along with OH and $NH_3$ (Bowers et al. 1983; Monnier et al 2000), this survey brings the total number of chemical species to 20 in this source, making it the most chemically complex circumstellar shell after IRC+10216, and comparable to CRL 618. Surprisingly, $^{32}SO_2$ and $^{34}SO_2$ are responsible for 35 of the total 130 observed emission lines. SiS is the next most prevalent species with 21 features, including isotopologues and vibrationally-excited states. Of the 18 species found here, three are completely new molecules - PO, AlO, and AlOH (Tenenbaum et al. 2007, Tenenbaum & Ziurys 2009, 2010), and four have never been seen before in O-rich shells (NaCl, PN, NS, and $HCO^+$). In addition, the line profiles in VY CMa exhibit complex velocity structures that cannot originate from a spherical circumstellar outflow, as found in IRC+10216 (see Ziurys et al. 2007, 2009).

## 4. Discussion

### *4.1 Carbon-Rich Versus Oxygen-Rich Circumstellar Chemistry*

This survey has enabled a direct comparison to be made between O-rich and C-rich stellar envelopes in the 1 mm band. It is obvious that free radicals are produced in both types of sources. In IRC+10216, nine radicals are found: CN, CP, SiC, CCH, l-$C_3H$, c-$C_3H$, $C_3N$, SiN, and $C_4H$. Five radical species are present in VY CMa (SO, CN, NS, PO, and AlO). Of the 32 different compounds present in IRC+10216, 23 contain carbon, or 72%. Many of these species are acetylenic chains ($C_2H$, $C_3H$, $C_4H$, $HC_3N$, and $SiC_2$). VY CMa has 6 carbon-bearing molecules

– 33% of the total. Nine molecules in VY CMa contain oxygen (~50% of the total), while only 4 have this element in IRC+10216 (~12%). It is clear that the C/O ratio does not completely control the relative chemistries in these two objects.

While species with any many as 7 atoms ($CH_3CCH$) were detected in IRC+10216, only diatomic and triatomic molecules are found in this spectral region in VY CMa. This result could simply be a question of sensitivity, or it could be chemistry. The more complex species in IRC+10216 contain carbon-carbon bonds, and are formed by neutral-neutral and ion-neutral reactions involving photodissociation products of "parent" species (Millar & Herbst 1994; Agundez et al. 2008). It is not clear that an equivalent chemistry exits involving oxygen chains. Partly hydrogenated oxygen chain molecules are known to be weakly bound (e.g. Mckay & Wright 1998; Murray et al. 2009), in contrast with the strong C-C bonds in carbon chain species. And while the rotational spectra of $HO_2$ (Saito 1977) and $HO_3$ (Suma et al. 2005) have been measured in the laboratory, there is no evidence for these molecules in the survey in VY CMa. Nevertheless, the bent-shaped radical $HO_3$ is a known intermediate species in the earth's atmosphere (Murray et al. 2009), and it may be detectable in VY CMa at lower frequencies.

Sulfur chemistry appears to play an important role in VY CMa, with 6 sulfur-bearing molecules observed, and the ubiquitous presence of $SO_2$ and SO. In contrast, only two silicon-containing molecules, SiO and SiS, were detected in this source. Sulfur's predominance may be a condensation effect. Three of the main dust constituents in VY CMa contain silicon (amorphous silicate, $Mg_2SiO_4$, $MgSiO_3$), while none include sulfur (Harwit et al. 2001). Therefore, excess sulfur may be available for gas phase molecule formation.

*4.2 Discovery of Circumstellar $CH_2NH$ in IRC+10216*

Nine favorable rotational transitions of methylenimine ($CH_2NH$) were detected in IRC+10216. The six unblended lines are shown in Figure 2 and line parameters are listed in Table 2. The features have consistent line shapes and intensities, with no obvious "missing" transitions. Three of these lines were also observed by He et al. (2008), but were marked as unidentified. $CH_2NH$ has previously been observed in dense clouds (Dickens et al. 1997) and in a galaxy (Salter et al. 2008), but never in a circumstellar envelope.

As Figure 2 illustrates, the $CH_2NH$ features appear slightly U-shaped, indicating that the molecule is present in the outer envelope. Rotational diagram analysis of the emission yields an excitation temperature of $T_{rot}$=26 K, consistent with outer envelope conditions, and a column density of $9\times10^{12}$ cm$^{-2}$, assuming a uniform filling factor. $CH_2NH$ is likely formed via gas-phase neutral-neutral reactions. One plausible sequence for creating $CH_2NH$ in IRC+10216 is the association of CH and $NH_3$, followed by rapid decomposition to $CH_2NH$ and H. Experiments have shown this scheme to have appreciable rate constants at temperatures as low as 23 K (Bocherel et al. 1996). One of the reactants, $NH_3$, has been detected in IRC+10216 (Monnier et al. 2000). The other reactant, CH, is a likely photodissociation product in the outer envelope.

Other partially saturated molecules have recently been detected in IRC+10216, such as $CH_2CN$, $CH_2CHCN$, and $CH_3CCH$, also arising from the outer envelope (Agundez et al. 2008). Column densities of these species fall in the range 6-18$\times10^{12}$ cm$^{-2}$, with excitation temperatures of 30-50 K, similar to the values found for $CH_2NH$. The detection of $CH_2NH$ is additional evidence that partially saturated hydrocarbons are formed in C-rich shells. The equivalent have yet to be found in O-rich environments, but the "inorganic" molecules $H_2SO_4$ or $ClO_2$ may be viable candidates.

*4.3 New Excited Vibrational States in VY CMa and IRC+10216*

Table 2 summarizes the vibrationally-excited states of CCH, NaCl, and $H_2O$ identified for the first time in an astronomical source. The presence of these transitions offers chemical insight into the two sources as well.

*a. The v=1 Level of NaCl in VY CMa*

In VY CMa, three rotational transitions of NaCl in the v=1 state were detected at 245 GHz, 271 GHz, and 284 GHz, with intensities ranging from 3-7 mK (Table 2). While ground state NaCl has been identified in a number of C-rich and O-rich circumstellar envelopes, including VY CMa (Milam et al. 2007, Highberger et al. 2003), the v=1 level is newly detected. The observed lines all exhibit widths of ~15-22 km $s^{-1}$ (see Table 2), similar to those of the v= 0 lines ($\Delta V_{1/2}$ = 13-24 km $s^{-1}$). Three other transitions of v=1 state lie in the survey region: the J=20→19 transition is completely obscured by a strong SO feature, the J=18→17 line is not observed due to insufficient sensitivity, and the J=17→16 line is contaminated by image features.

The v=1 level in NaCl lies ~520 K above the ground state, and therefore it is striking that lines originating in this state are only 2 to 3 times less intense than their v=0 counterparts. The estimated vibrational temperature is $T_{vib}$~ 430 K. NaCl is likely formed near the stellar photosphere (Milam et al. 2007), where the gas is both hot (T~3000 K) and dense, and may cause vibrational excitation. Radiative pumping at 3 μm could also contribute to populating the v=1 level. In contrast, the v=1 lines of NaCl in IRC+10216 were not observed down to upper limits of ~4-5 mK (2 MHz resolution), although the v=0 transitions had intensities of ~20 mK. The presence of NaCl in both sources indicates that LTE chemistry is occurring near the stellar photosphere (Tsuji 1973, Ziurys 2006); the v=1 level suggests that this chemistry is taking place at higher temperatures in VY CMa.

*b. Vibrationally Excited CCH ($v_2$=1) in IRC+10216*

Three rotational transitions originating in the $v_2$=1 bending mode of CCH ($X^2\Sigma^+$), located 530 K above ground state, were observed toward IRC+10216. These lines are the vibrational satellites of the N=3→2 transition, which form a doublet of doublets from l-type and spin-rotation interactions. (The proton hyperfine splitting is negligible). The fourth line at 261.2 GHz is obscured by an $SiC_2$ $v_3$=1 feature. Ground state CCH has been known to exist in space since 1974, but past searches for vibrationally-excited CCH have been unsuccessful (e.g. Woodward et al. 1987).

The detected features appear slightly U-shaped with the typical line width of ~28 km $s^{-1}$ (see Figure 3), indicating that the $v_2$=1 level of CCH must originate in the outer shell. Since thermal or radiative excitation to the $v_2$=1 level is improbable in the cool outer envelope, it is likely that the photodissociation of acetelyne ($C_2H_2$) directly creates vibrationally-excited CCH. Laboratory studies have shown that UV photodissociation of $C_2H_2$ favors formation of CCH in the $v_2$ bending mode (Morduant et al. 1998) or in the $A^2\Pi$ electronic state (Lai et al. 1996). The rotational spectrum of $^1A^2\Pi$ CCH has not been measured, but this state could contribute to the production of the $v_2$ bending mode.

*c. A Remarkable New Water Maser: $H_2O$ ($v_2$=2) in VY CMa*

The most intense emission line observed in the 1 mm band of VY CMa is the $T_A$*~14 K feature arising from the $v_2$=2, $J_{Ka,Kc} = 6_{5,2} \rightarrow 7_{4,3}$ transition of water at 268,149 MHz, displayed in Figure 3. Water has never been seen before in the $v_2$=2 state which has an $E_u$ ~6040 K, though there have been many reports of lines in the $v_2$=1 state (e.g. Menten et al. 2006). The width of the line, $\Delta V_{1/2}$=4 km $s^{-1}$, is more narrow than typical VY CMa emission features. Observations over three epochs in 2008 (April-June) showed no variability in line intensity, and comparison of the orthogonal polarization data did not reveal significant differences. The substantial intensity of the line and the narrow width indicate maser action, and potential use for mm-wave VLBI. Two other water lines, both in the $v_2$=1 state, were detected in the VY CMa data: the $J_{Ka,Kc}=5_{5,0}\rightarrow6_{4,3}$ (232,686 MHz) and the $J_{Ka, Kc}=7_{0,7}\rightarrow8_{6,3}$ (263,451 MHz) transitions. The intensities of these features are $T_A$*=0.02-0.04 K, and the emission appears to be broader with multiple peaks. Both of these lines have been previously observed in circumstellar envelopes (Menten & Melnick 1989; Alcolea & Menten 1993). The $v_2$=1 state has also been tentatively observed in the far-IR (Polehampton et al. 2010). These data are further evidence that in supergiants/hypergiants oxygen is primarily contained in $H_2O$. Therefore, the abundance of CO is low (Ziurys et al. 2009, Dinh-v-Trung et al. 2009) and there is carbon remaining to form other molecules. Despite the chemical richness of IRC+10216, oxygen rich envelopes such as that of VY CMa offer a new perspective on circumstellar chemistry.

**Acknowledgements**: This research is supported by NSF Grant AST-06-07803 and AST-09-06534 E.D.T. acknowledges support from an NSF graduate research fellowship.

**References**

Agúndez, M. & Cernicharo, J. 2006, ApJ, 650, 374

Agúndez, M., Fronfria, J. P., Cernicharo, J., Pardo, J. R. & Guélin, M. 2008, A&A, 479, 493

Alcolea, J. & Menten, K. M. 1993, in Astrophysical Masers, eds. A. W. Clegg & G. E. Nedoluha (Berlin: Springer), 399

Avery, L. W. et al. 1992, ApJS, 83, 363

Bocherel, P., Herbert, L. B., Rowe, B. R., Sims, I. R., Smith, I. W. M., & Travers, D. 1996, J. Phys. Chem., 100, 30

Bowers, P. E., Johnston, K. J. & Spencer, J. H. 1983, ApJ, 274, 733

Caris, M., Lewen, F., & Winnewisser, G. 2002, Z. Naturforsch, 57 a, 663

Cernicharo, J., Guélin, M. & Kahane, C. 2000, A&AS, 142, 181

Dickens, J. E., Irvine, W. M., De Vries, C. H. & Ohishi, M. 1997, ApJ, 479, 307

Dinh-V-Trung, Muller, S., Lim, J., Kwok, S. & Muthu, C. 2009, 697, 409

Dodd, J. L., Tenenbaum, E. D., Woolf, N. J., Ziurys, L. M. in preparation (Paper III)

He, J. H., Dinh-V-Trung, Kwok, S., Müller, H. S. P., Zhang, Y., Hasegawa, T., Peng, T. C. & Huang, Y. C. 2008, ApJS, 177, 275

Harwit, M., Malfait, K., Decin, L., Waelkens, C., Feuchtgruber, H. & Melnick, G. J. 2001, ApJ, 557, 844

Herwig, F. 2005, ARA&A, 43, 435

Highberger, J. L., Thompson, K. J., Young, P. A., Arnett, D. & Ziurys, L. M. 2003, ApJ, 593, 393

Kawaguchi, K., Kasai, Y., Ishikawa, S., Kaifu, N. 1995, PASJ, 47, 853

Kwok, S. 2004, Nature, 430, 985

Lai, L. –H., Che, D. -C. & Liu, K. 1996, J. Phys. Chem. 100, 6376

Mckay, D. J. & Wright, J. S. 1998, J. Am. Chem. Soc., 120, 1003

Menten, K. M. & Melnick, G. J. 1989, ApJ, 341, L91

Menten, K. M., Phillips, S. D., Güsten, R., Alcolea, J., Polehampton, E. T. & Brünken, S. 2006, A&A, 454, L107

Milam, S. N., Apponi, A. J., Woolf, N. J., & Ziurys, L. M. 2007, ApJ, 668, L131

Milam, S. N., Woolf, N. J. & Ziurys, L. M. 2009, ApJ, 690, 837

Millar, T. J. & Herbst, E. 1994, A&A, 288, 561

Monnier, J. D., Danchi, W. C., Hale, D. S., Lipman, E. A., Tuthill, P. G., & Townes, C. H. 2000, ApJ, 543, 861

Morduant, D. H., Ashfold, M. N. R., Dixon, R. N., Löffler, P., Schnieder, L. & Welge, K. H. 1998, J. Chem. Phys. 108, 519

Murray, C., Derro, E. L., Sechler, T. D. & Lester, M. I. 2009, Acc. Chem. Res. 42, 419

Omont, A., Lucas, R., Morris, M. & Guilloteau, S. 1993, A&A, 267, 490

Olofsson, H., Lindqvist, M., Winnberg, A., Nyman, L. –A. & Nguyen-Q-Rieu 1991, A&A, 245, 611

Olofsson, H. 2005, in Proc. of the Dusty Molecular Universe: a Prelude to Herschel and ALMA, ed. A. Wilson (Noordwijk: ESA), 223

Pardo, J. R., Cernicharo, J., Goicoechea, J. R., Guélin, M. & Asensio Ramos, A. 2007, ApJ, 661, 250

Pearson, J. C., Anderson, T., Herbst, E., De Lucia, F. C., & Helminger, P. 1991, ApJ, 379, L41

Polehampton, E. D., Menten, K. M., van der Tak, F. F. S. & White, G. J.

Saito, S. 1977, J. Mol. Spectrosc. 65, 229

Salter, C. J., Ghosh, T., Catinella, B., Lebron, M., Lerner, M. S., Minchin, R. & Momjian, E. 2008, AJ, 136, 389

Smith, N., Humphreys, R. M., Davidson, K., Gehrz, R. D., Schuster, M. T., & Krautter, J. 2001, AJ, 121, 1111

Suma, K. Sumiyoshi, Y. & Endo, Y. 2005, Science, 308, 1885

Tenenbaum, E. D., Woolf, N. J., & Ziurys, L. M. 2007, ApJ, 666, L29

Tenenbaum, E. D. & Ziurys, L. M. 2008, ApJ, 680, L121

Tenenbaum, E. D. & Ziurys, L. M. 2009, ApJ, 694, L59

Tenenbaum. E. D. & Ziurys, 2010, ApJ, 712, L93

Tenenbaum, E. D., Dodd, J. L., Milam, S. N., Woolf. N. J. & Ziurys, L. M. 2010 ApJS, in press (Paper II)

Tsuji, T. 1973, A&A, 23, 411

Willacy, K. & Millar, T. J. 1997, A&A, 324, 237

Wilson, R. W., Solomon, P. M., Penzias, A. A. & Jefferts, K. B. 1971, ApJ, 169, L35

Woodward, D. R., Pearson, J. C., Gottlieb, C. A., Guélin, M. & Thaddeus, P. 1987, A&A, 186, L14

Ziurys, L. M. 2006, PNAS, 103, 12274

Ziurys, L. M., Milam, S. N., Apponi, A. J., & Woolf, N. J. 2007, Nature, 447, 1094

Ziurys, L. M., Tenenbaum, E. D., Pulliam, R. L., Woolf, N. J., & Milam, S. N. 2009, ApJ, 695, 1604

**Figure Captions**

Figure 1: The complete ARO SMT spectral survey of VY CMa and IRC+10216 (214.5-285.5 GHz). The intensity scale is the same for both sources, and is truncated at $T_A^*$=0.1 K to show the weaker lines. The lower panel displays a 1 GHz section of the survey, centered at 267 GHz (smoothed to a resolution of 2 MHz per channel). The panel highlights molecular features common to both sources: HCN $v_2=1^{1d}$ (J=3→2), $^{29}$SiS (J=15→14), Na$^{37}$Cl (J=21→20), and $HCO^+$ (J=3→2). The IRC+10216 spectrum also contains $PH_3$, and various other vibrationally excited HCN lines that are notably missing from VY CMa. Assumed velocities are $V_{LSR}$= -26 km $s^{-1}$ (IRC+10216) and $V_{LSR}$=19 km $s^{-1}$ (VY CMa).

Figure 2: Spectra of the six unblended $CH_2NH$ emission features observed in the survey toward IRC +10216. The lines appear slightly U-shaped, indicating a shell-like distribution of this molecule in the circumstellar envelope. These data have been smoothed to 2 MHz resolution.

Figure 3: Spectra showing the detections of new vibrationally-excited states in the survey data. The top panel displays a new water maser in VY CMa originating in the $v_2$=2 level, $J_{Ka,Kc}=6_{5,2}\rightarrow7_{4,3}$. The second panel shows the J=19→18 transition of NaCl in the v=1 level towards VY CMa, which exhibits a very narrow linewidth. The third and fourth panels display two lines of CCH in the $v_2$=1 mode at 260,448 MHz and 259,152 MHz, respectively, observed towards IRC+10216. The vibrationally-excited CCH lines are U-shaped, indicating an origin in the outer shell. Spectral resolution in the top panel is 1 MHz, and in the other three panels the data have been smoothed to 2 MHz resolution. The $\Delta V_{LSR}$ scale is plotted with respect to $V_{LSR}$= -26 km $s^{-1}$ for IRC+10216 and $V_{LSR}$=19 km $s^{-1}$ for VY CMa.

**Table 1. Species Observed in this Survey Toward IRC+10216 and VY CMa [a)]**

| IRC +10216 | | VY CMa | |
|---|---|---|---|
| **Molecule** | **# Lines** | **Molecule** | **# Lines** |
| CO | 5 | CO | 2 |
| SiO | 6 | SiO | 12 |
| SiS | 47 | SiS | 21 |
| CS | 10 | CS | 1 |
| CN | 16 | CN | 3 |
| HCN | 17 | HCN | 3 |
| HNC | 3 | HNC | 1 |
| NaCl | 11 | NaCl | 11 |
| PN | 2 | PN | 2 |
| $HCO^+$ | 1 | $HCO^+$ | 1 |
| $PH_3$* | 1 | NS | 2 |
| $CH_2NH$ | 9 | PO* | 6 |
| CP | 2 | AlO* | 2 |
| SiC | 8 | AlOH* | 2 |
| AlCl | 10 | SO | 7 |
| KCl | 9 | $H_2O$ | 3 |
| AlF | 2 | $SO_2$ | 36 |
| SiN | 4 | $H_2S$ | 1 |
| HCP | 2 | U | 14 |
| $SiC_2$ | 138 | | |
| CCH | 17 | | |
| NaCN | 38 | | |
| l-$C_3H$ | 17 | | |
| c-$C_3H$ | 13 | | |
| $C_3N$ | 13 | | |
| $H_2CO$ | 3 | | |
| $H_2CS$ | 9 | | |
| $HC_3N$ | 13 | | |
| c-$C_3H_2$ | 16 | | |
| $C_4H$ | 124 | | |
| $CH_3CN$ | 24 | | |
| $CH_3CCH$ | 7 | | |
| U | 123 | | |
| Total | 720 | Total | 130 |

[a)] Gray background indicates molecules common to both sources
* New species detected in this survey

**Table 2. Line Parameters of New Circumstellar Molecules and Vibrationally Excited States**

| Molecule | Star | Transition | $\nu_{rest}$ (MHz) | $T_A^*$ (K) | $\int T_A^* dV$ (K km s$^{-1}$) | $\Delta V_{1/2}$ km s$^{-1}$ |
|---|---|---|---|---|---|---|
| $H_2O$, $v_2 = 2$ | VY CMa | $J_{Ka, Kc} = 6_{5, 2} \rightarrow 7_{4, 3}$ | 268,149.1[a] | 14.030 ± 0.001 | 54.917 ± 0.053 | 3.7 ± 1.1 |
| NaCl, v = 1 | VY CMa | J = 19 → 18 | 245,401.1[b] | 0.003 ± 0.001 | 0.046 ± 0.025 | 22.0 ± 2.4 |
| | | J = 21 → 20 | 271,170.1[b] | 0.005 ± 0.002 | 0.070 ± 0.019 | 15.4 ± 4.4 |
| | | J = 22 → 21 | 284,047.6[b] | 0.007 ± 0.002 | 0.088 ± 0.027 | 16.8 ± 4.2 |
| CCH, $v_2 = 1$ | IRC +10216 | N = 3 → 2  J = 5/2 → 3/2 [e] | 259,152.2[c] | 0.005 ± 0.002 | 0.073 ± 0.030 | 27.8 ± 2.3 |
| | | J = 7/2 → 5/2 [e] | 260,448.0[c] | 0.007 ± 0.002 | 0.095 ± 0.037 | 29.9 ± 2.3 |
| | | J = 7/2 → 5/2 [f] | 262,488.6[c] | 0.005 ± 0.002 | 0.039 ± 0.025 | 25.1 ± 4.6 |
| $CH_2NH$ | IRC +10216 | $J_{Ka, Kc} = 1_{1, 1} \rightarrow 0_{0, 0}$ | 225,554.7 | 0.013 ± 0.003 | 0.411 ± 0.040 | 29.2 ± 2.7 |
| | | $J_{Ka, Kc} = 6_{1, 5} \rightarrow 6_{0, 6}$ | 226,548.8 | 0.009 ± 0.002 | 0.113 ± 0.040 | 26.5 ± 2.6 |
| | | $J_{Ka, Kc} = 4_{1, 4} \rightarrow 3_{1, 3}$ | 245,126.0 | 0.012 ± 0.003 | 0.356 ± 0.035 | 26.9 ± 2.4 |
| | | $J_{Ka, Kc} = 6_{0, 6} \rightarrow 5_{1, 5}$ [d] | 251,421.4 | ~ 0.008 | — | — |
| | | $J_{Ka, Kc} = 4_{0, 4} \rightarrow 3_{0, 3}$ [d] | 254,685.2 | ~ 0.021 | — | — |
| | | $J_{Ka, Kc} = 4_{2, 3} \rightarrow 3_{2, 2}$ [d] | 255,840.4 | ~ 0.008 | — | — |
| | | $J_{Ka, Kc} = 4_{2, 2} \rightarrow 3_{2, 1}$ | 257,112.8 | 0.005 ± 0.002 | 0.111 ± 0.070 | 25.7 ± 4.7 |
| | | $J_{Ka, Kc} = 4_{1, 3} \rightarrow 3_{1, 2}$ | 266,270.1 | 0.019 ± 0.001 | 0.561 ± 0.045 | 27.0 ± 2.3 |
| | | $J_{Ka, Kc} = 2_{1, 2} \rightarrow 1_{0, 1}$ | 284,254.1 | 0.010 ± 0.002 | 0.326 ± 0.018 | 27.4 ± 2.1 |

[a] Pearson et al. 1991
[b] Caris et al. 2002
[c] Woodward et al. 1987
[d] Blended with another emission line

**Figure 1.**

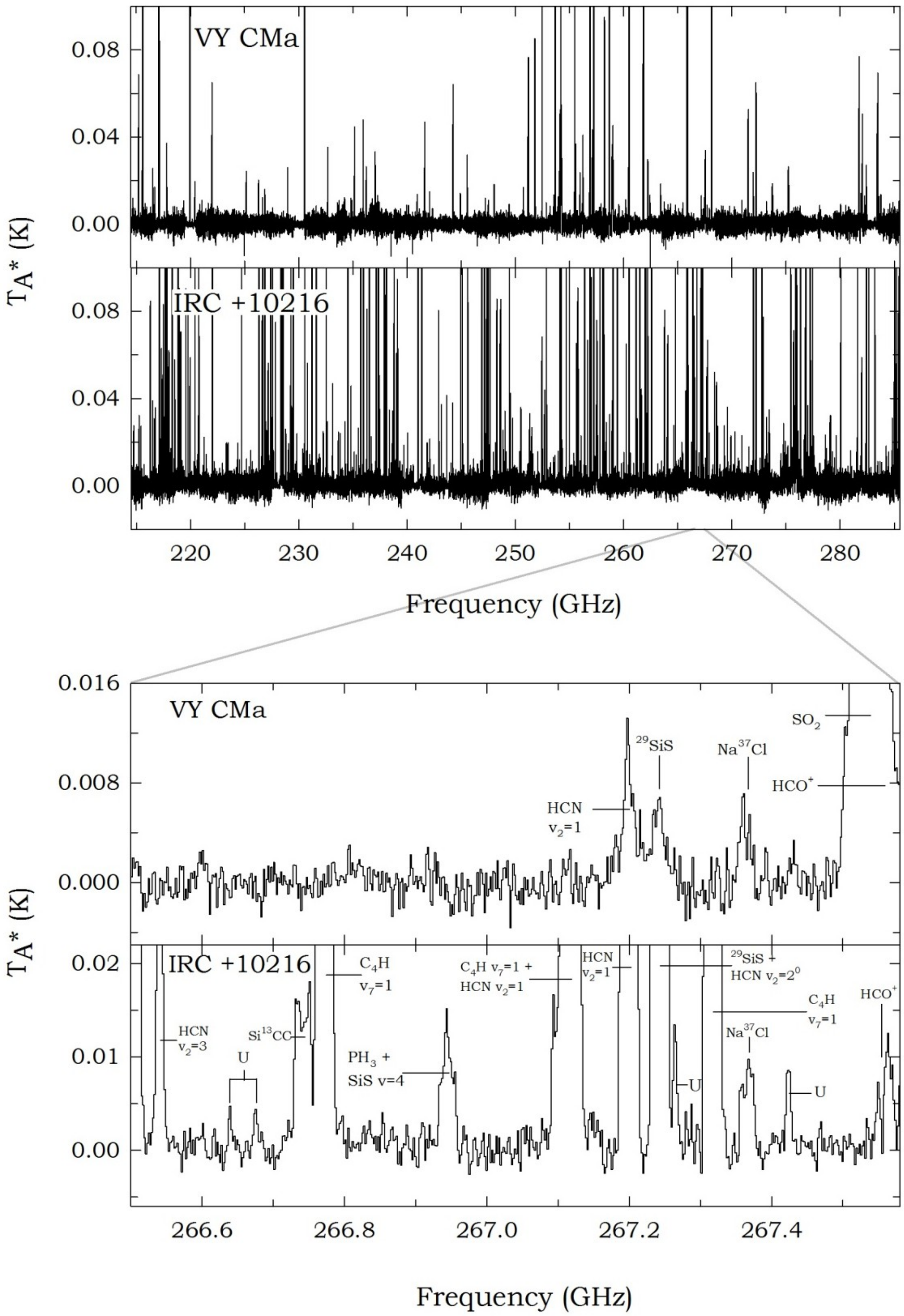

VY CMa
IRC +10216
TA* (K)
Frequency (GHz)
HCN v2=1
29SiS
Na37Cl
SO2
HCO+
HCN v2=3
U
Si13CC
C4H v7=1
PH3 + SiS v=4
C4H v7=1 + HCN v2=1
29SiS + HCN v2=20

Figure 2.

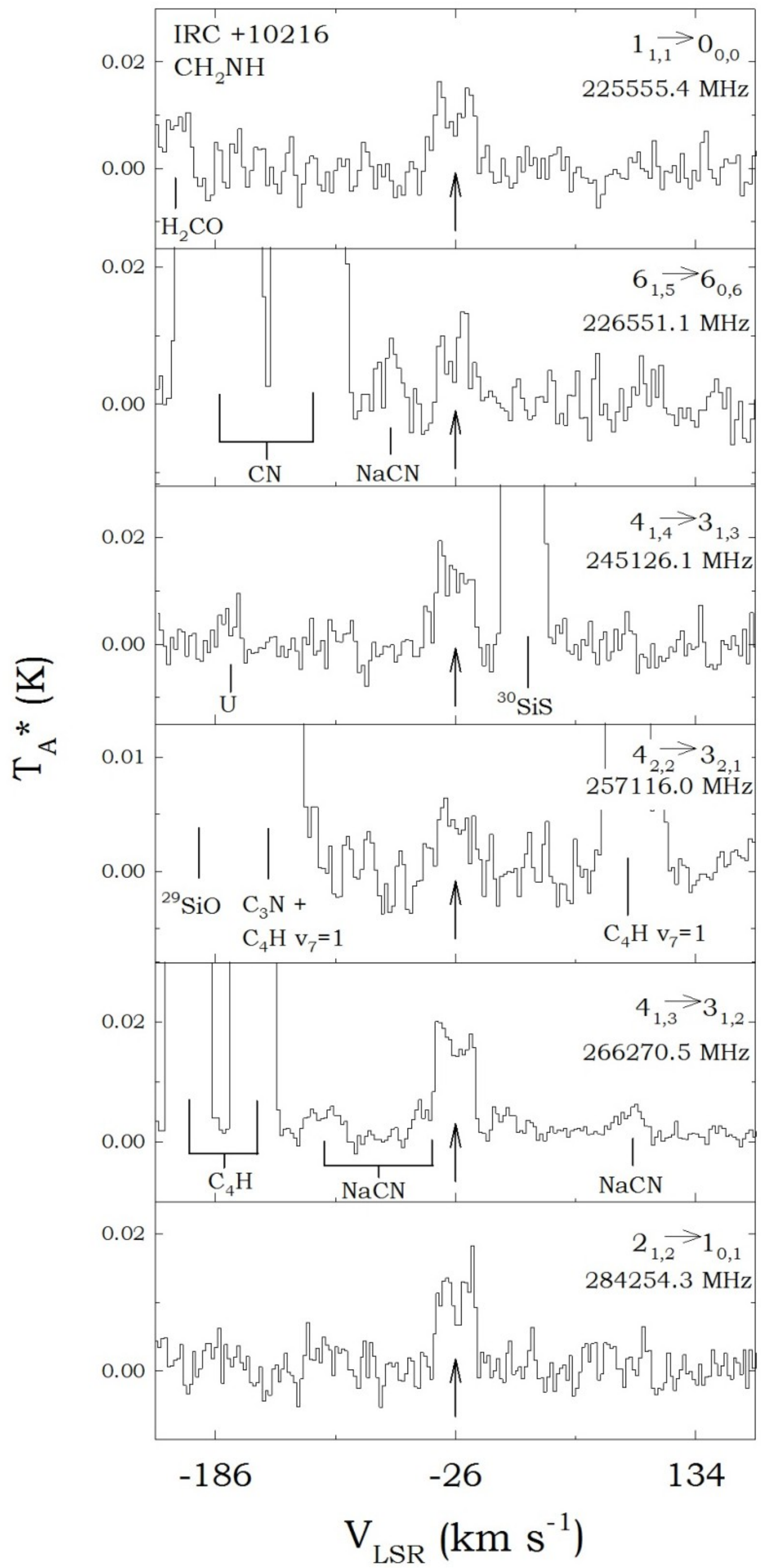
IRC +10216
CH$_2$NH
$1_{1,1} \rightarrow 0_{0,0}$
225555.4 MHz
H$_2$CO
$6_{1,5} \rightarrow 6_{0,6}$
226551.1 MHz
CN
NaCN
$4_{1,4} \rightarrow 3_{1,3}$
245126.1 MHz
U
$^{30}$SiS
$4_{2,2} \rightarrow 3_{2,1}$
257116.0 MHz
$^{29}$SiO
C$_3$N +
C$_4$H v$_7$=1
C$_4$H v$_7$=1
$4_{1,3} \rightarrow 3_{1,2}$
266270.5 MHz
C$_4$H
NaCN
NaCN
$2_{1,2} \rightarrow 1_{0,1}$
284254.3 MHz
T$_A$* (K)
V$_{LSR}$ (km s$^{-1}$)

Figure 3.

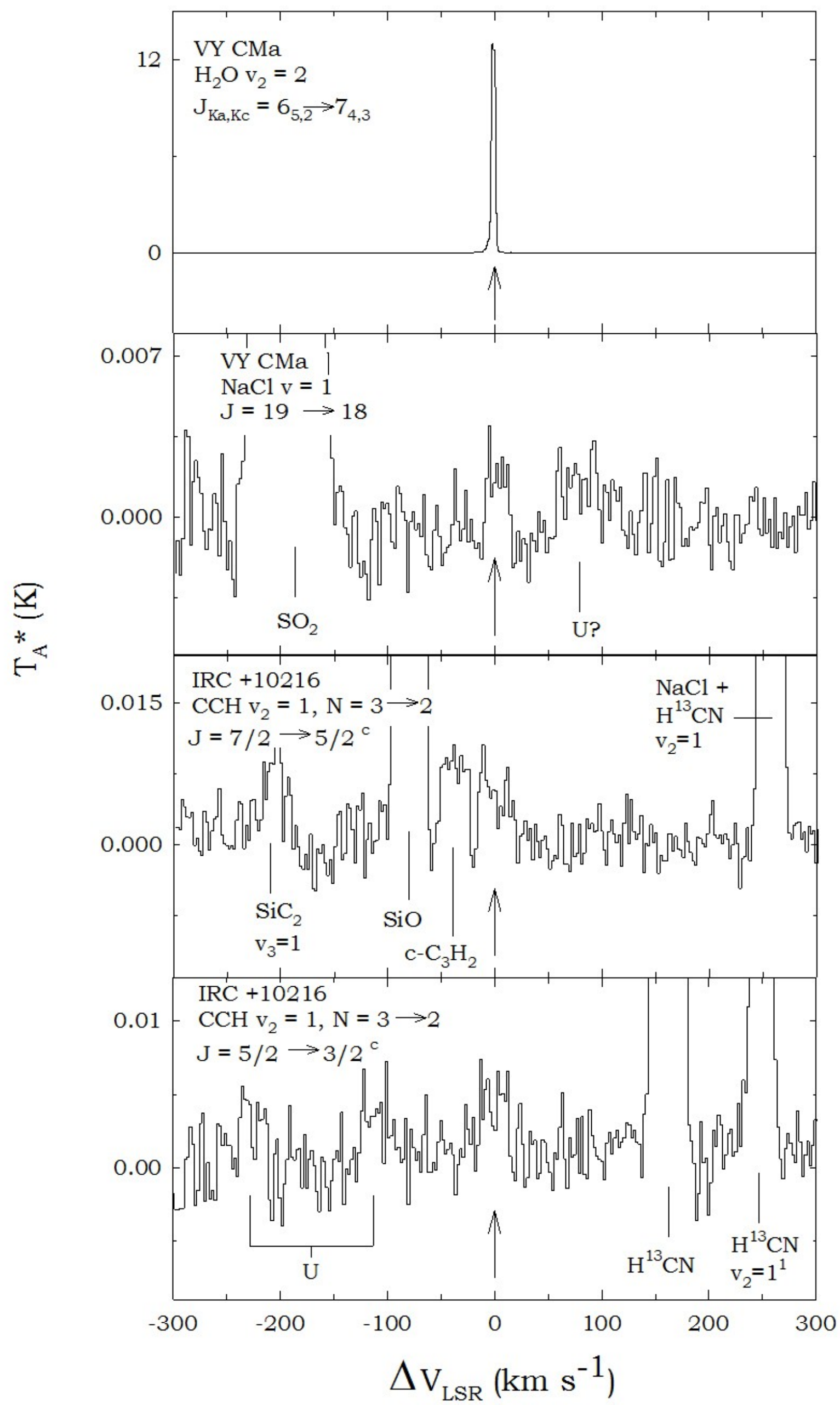

VY CMa
$H_2O$ $v_2$ = 2
$J_{Ka,Kc} = 6_{5,2} \rightarrow 7_{4,3}$
12
0
VY CMa
NaCl v = 1
J = 19 → 18
0.007
0.000
$SO_2$
U?
IRC +10216
CCH $v_2$ = 1, N = 3 → 2
J = 7/2 → 5/2 c
NaCl +
$H^{13}CN$
$v_2$=1
0.015
0.000
$SiC_2$
$v_3$=1
SiO
c-$C_3H_2$
IRC +10216
CCH $v_2$ = 1, N = 3 → 2
J = 5/2 → 3/2 c
0.01
0.00
U
$H^{13}CN$
$H^{13}CN$
$v_2=1^1$
$T_A^*$ (K)
-300
-200
-100
0
100
200
300
$\Delta V_{LSR}$ (km $s^{-1}$)